\documentclass{iopart}
\usepackage{graphicx}
\usepackage{iopams}

\begin{document}

\title[]{Properties of Bose Gases with Raman-Induced Spin-Orbit Coupling}
\author{Wei Zheng$^1$, Zeng-Qiang Yu$^{1,2}$, Xiaoling Cui$^1$ and Hui Zhai$%
^1$}
\address{$^1$ Institute for Advanced Study, Tsinghua University, Beijing,
100084, China.\\ $^2$ Dipartimento di Fisica, Universit\`{a} di Trento and
INO-CNR BEC Center, I-38123 Povo, Italy.}

\begin{abstract}
In this paper we investigate the properties of Bose gases with Raman-induced
 spin-orbit(SO) coupling. It is found that the SO coupling can greatly modify the single particle density-of-state, and thus lead to non-monotonic behavior of the condensate depletion, the Lee-Huang-Yang correction of ground-state energy and the transition temperature of a non-interacting Bose-Einstein condensate. The presence of the SO coupling also breaks the Galileaan invariance, and this gives two different critical velocities, corresponding to the movement of the condensate and the impurity respectively. Finally, we show that with SO coupling, the interactions modify the BEC transition temperature even at Hartree-Fock level, in contrast to the ordinary Bose gas without SO coupling. All results presented here can be directly verified in the current cold atom experiments using Raman laser-induced gauge field.
\end{abstract}

\maketitle

\section{Introduction}

Recently one of the major progresses in cold atom physics is the realization
of spin-orbit (SO) coupling for neutral atoms with the idea of light induced
synthetic gauge field\cite%
{NIST_Nature,Zhang_Jing_fermion,MIT_fermion,USTC_boson}. Previously, effects
of spin-orbit coupling have only been studied in fermionic matters such as
electron gases, while this progress opens up the new opportunity of studying
SO coupling in bosonic systems. This investment is interesting because SO
coupling changes many basic properties of Bose-Einstein condensation (BEC)
in both quantitative and qualitative way.

Although SO coupling emerges at
single particle level, it may have a significant effect to the many-body
behavior due to the modification of the single particle density-of-state
(DoS). One particular example is the Rashba SO coupled Bose gases. Because
of the dramatic change of low energy DoS, many intriguing many-body
phenomena have been predicted there, and some questions still have not been
thoroughly understood yet. While a number of theoretical works have studied
its low-temperature properties, including ground state phases, quantum and
thermal fluctuations, Bose condensation transition and superfluiditiy \cite%
{Zhai,Victor,Wu,Gordom,Cui,Qi,Biao_Wu}, the experimental realization of
Rashba SO coupling remains to be a great challenge. On the other hand, the
type of equal Rashba and Dresselhaus SO coupling has been realized in
current experiment with two photon Raman transition. Although the change of
low-energy DoS is not as dramatic as Rashba case, such a SO coupling
leads to a non-monotonic behavior of low-energy DoS as a function of Raman
strength, which should result in nontrivial manifestation of many-body
properties. So far, a few papers have studied the ground state phase
diagram, collective modes and response function for such SO coupled
condensate \cite{Ho, Stringari1,Stringari2,Stringari3,Chen,Zheng}, but there
are still several fundamental properties which have not been explored, in
particular, the depletion of the condensate, the BEC transition temperature and
the superfluid critical velocities. In this work we report our theoretical
studies of these properties.

\section{Single particle Hamiltonian and mean-field phase diagram}

\subsection{Single particle Hamiltonian}

In current experiments, two counter-propagating Raman beams are applied to
couple two hyperfine levels of an alkali atom \cite%
{NIST_Nature,Zhang_Jing_fermion,MIT_fermion,USTC_boson}, which gives rise to
the single-particle Hamiltonian as \cite{Zhai_review} (set $\hbar =1$)
\begin{equation}
\hat{H}_{0}=\left(
\begin{array}{cc}
\left( \mathbf{\hat{p}}-k_{r}\mathbf{e}_{x}\right) ^{2}/2m+\delta /2 &
\Omega /2 \\
\Omega /2 & \left( \mathbf{\hat{p}}+k_{r}\mathbf{e}_{x}\right)
^{2}/2m-\delta /2%
\end{array}%
\right)  \label{H0}
\end{equation}%
where $k_{r}$ is the recoil momentum, $\Omega $ is the Raman coupling
strength, and $\delta $ is the two-photon detunning. With Pauli matrix, the
Hamiltonian in Eq. (\ref{H0}) can be rewritten as
\begin{equation}
\hat{H}_{0}=\frac{\mathbf{\hat{p}}^{2}}{2m}-\mathbf{B}_{\mathbf{p}}\cdot
\boldsymbol{\sigma }+E_r
\end{equation}%
where $E_r=k_r^2/(2m)$ is the recoil energy, and $\mathbf{B}_{\mathbf{p}}=(-\Omega /2,0,k_{r}p_{x}/m-\delta /2)$ depends
on momentum $p_{x}$ and yields the locking between spin and momentum. This
Hamiltonian preserves spatial translational symmetry, and momentum $\mathbf{p%
}$ is a good quantum number. Another quantum number of this Hamiltonian is
"helicity" $h=\pm $, which denotes for $\mathbf{B}_{\mathbf{p}}$ parallel or
anti-parallel to spin $\boldsymbol{\sigma }$. Thus, the eigen-energies of
two helicity branches are given by
\begin{equation}
\varepsilon _{\mathbf{p},\pm }=\frac{p^{2}}{2m}+\frac{k_{r}^{2}}{2m}\pm
\sqrt{\left( \frac{p_{x}k_{r}}{m}-\frac{\delta }{2}\right) ^{2}+\left( \frac{%
\Omega }{2}\right) ^{2}}
\end{equation}%
and their wave-functions are given by
\begin{equation}
\phi _{\mathbf{p},+}(\mathbf{r})=e^{i\mathbf{p}\cdot \mathbf{r}}\left(
\begin{array}{c}
\sin \theta _{\mathbf{p}} \\
\cos \theta _{\mathbf{p}}%
\end{array}%
\right) ;\quad \phi _{\mathbf{p},-}(\mathbf{r})=e^{i\mathbf{p}\cdot \mathbf{r}}\left(
\begin{array}{c}
\cos \theta _{\mathbf{p}} \\
-\sin \theta _{\mathbf{p}}%
\end{array}%
\right)
\end{equation}%
with
\[
\theta _{\mathbf{p}}=\arcsin\left[ \frac{1}{2}%
\left( 1+\frac{p_{x}k_{r}/m-\delta /2}{\sqrt{\left(
p_{x}k_{r}/m-\delta /2\right) ^{2}+\Omega ^{2}/4}}\right) \right]^{1/2}
\label{sincos}
\] %
%
%
%
%
%
%
With the spectrum, the single particle DoS can be calculated
straightforwardly as
\begin{equation}
D(\epsilon )={\frac{1}{V}}\sum_{\mathbf{p}}\big[\delta _{\mathrm{D}%
}(\varepsilon -\varepsilon _{\mathbf{p,+}})+\delta _{\mathrm{D}}(\varepsilon
-\varepsilon _{\mathbf{p,-}})\big]
\end{equation}%
where $\delta _{\mathrm{D}}$ is Dirac delta function. %

\begin{figure}[tbp]
\includegraphics[width=5 in]{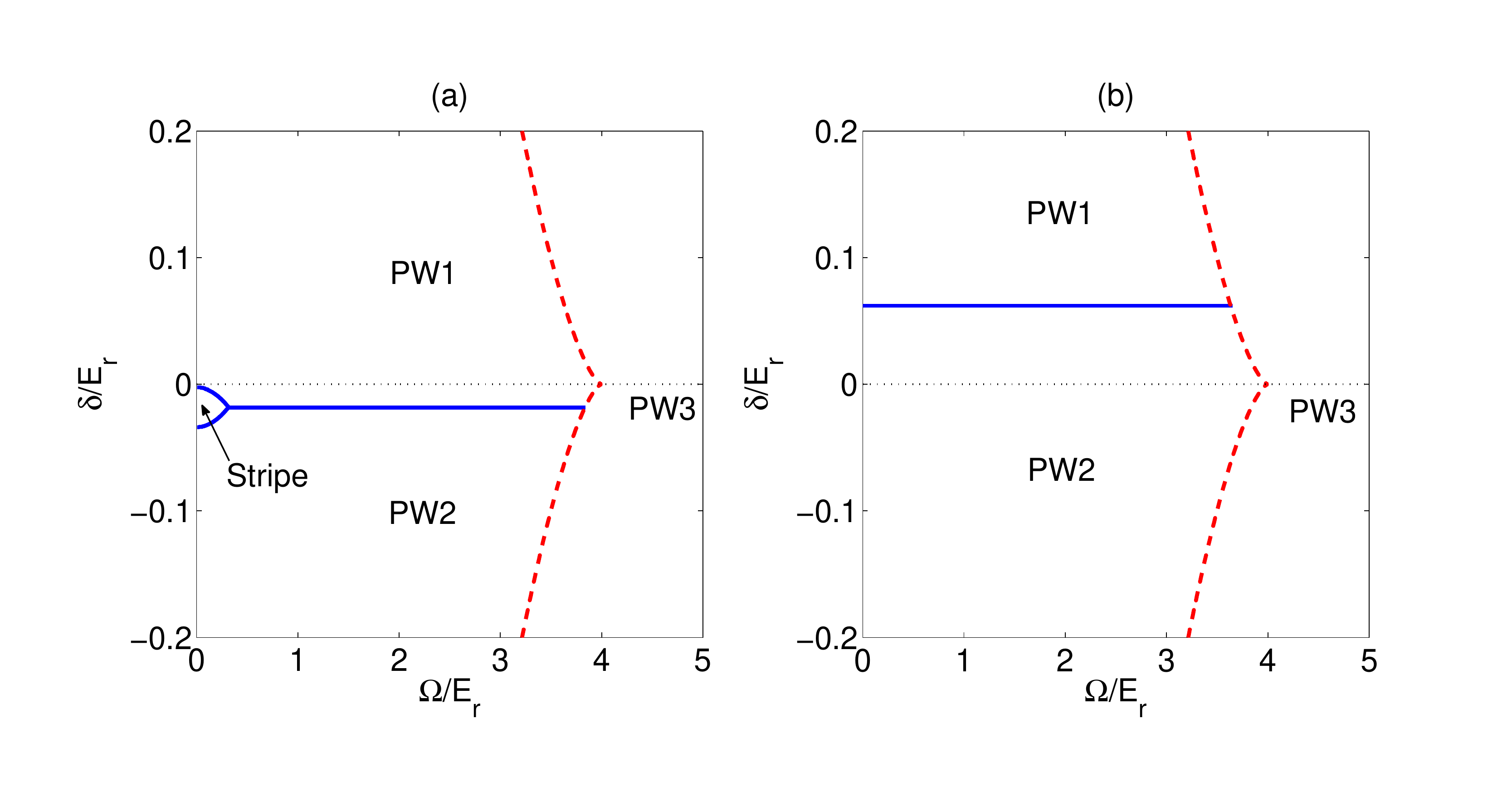}
\caption{The phase diagrams of $^{87}$Rb (a) and $^{23}$Na (b).
The two pseudo-spin states are $m=0,-1$ states of $F=1$ hyperfine states.
On the right of red dashed line, the single particle spectrum has one local minimum, and there is only one plane wave phase (PW3).
On the left of red dashed line, the single particle spectrum has two local minima, and the horizontal
blue line separates two plane-wave condensates at different momenta (PW1 and PW2). Within the region rounded by two blues lines it is the stripe phase, in which the condensate coherently occupies two different momenta.
Here $E_{r}=2 \protect\pi \times 2.2\mathrm{kHz}$.}
\end{figure}

\subsection{Mean-field phase diagram}

The interaction of a two-component Bose gas is generally written as
\begin{equation}
\hat{H}_{I}=\frac{1}{2}\int d^{3}\mathbf{r}\left( g_{\uparrow \uparrow }\hat{%
\psi}_{\uparrow }^{\dag }\hat{\psi}_{\uparrow }^{\dag }\hat{\psi}_{\uparrow }%
\hat{\psi}_{\uparrow }+g_{\downarrow \downarrow }\hat{\psi}_{\downarrow
}^{\dag }\hat{\psi}_{\downarrow }^{\dag }\hat{\psi}_{\uparrow }\hat{\psi}%
_{\downarrow }+2g_{\uparrow \downarrow }\hat{\psi}_{\uparrow }^{\dag }\hat{%
\psi}_{\downarrow }^{\dag }\hat{\psi}_{\downarrow }\hat{\psi}_{\uparrow
}\right) .
\end{equation}%
With mean-field approximation, the interaction energy is given by
\begin{equation}
\mathcal{E}_{I}=\frac{1}{2}\int d^{3}\mathbf{r}\left( g_{\uparrow \uparrow
}n_{\uparrow }^{2}(\mathbf{r})+g_{\downarrow \downarrow }n_{\downarrow }^{2}(%
\mathbf{r})+2g_{\uparrow \downarrow }n_{\uparrow }(\mathbf{r})n_{\downarrow
}(\mathbf{r})\right)  \label{int}
\end{equation}%
When $\delta $ and $\Omega $ are both small, $\epsilon _{-}$ has two local
minima located at $\mathbf{k}_{\pm }$. Without loss of generality, we can
assume the condensate wave function as

\begin{equation}
\varphi(\mathbf{r}) =\sqrt{n_{0}}\left[ \cos \alpha \left(
\begin{array}{c}
\cos \theta _{+} \\
-\sin \theta _{+}%
\end{array}%
\right) e^{ i\mathbf{p}_{+} \cdot \mathbf{r}} +\sin \alpha \left(
\begin{array}{c}
\sin \theta _{-} \\
-\cos \theta _{-}%
\end{array}%
\right) e^{ i\mathbf{p}_{-} \cdot \mathbf{r}} \right] ,  \label{vari}
\end{equation}%
where $0\leq\alpha\leq\pi /2$, $\cos \alpha$ and $\sin \alpha$ are superposition coefficients. It gives
\begin{eqnarray}
n_{\uparrow } &=&n_{0}\left[ \cos ^{2}\alpha \cos ^{2}\theta _{+}+\sin
^{2}\alpha \cos ^{2}\theta _{-}+\sin 2\alpha \cos \theta _{+}\cos \theta
_{-}\cos (\delta \mathbf{p}\cdot \mathbf{r})\right] ,  \nonumber \\
n_{\downarrow } &=&n_{0}\left[ \cos ^{2}\alpha \sin ^{2}\theta _{+}+\sin
^{2}\alpha \sin ^{2}\theta _{-}+\sin 2\alpha \sin \theta _{+}\sin \theta _{%
\mathbf{-}}\cos (\delta \mathbf{p}\cdot \mathbf{r})\right] ,  \label{density}
\end{eqnarray}%
where $\delta \mathbf{p}=\mathbf{p}_{+}-\mathbf{p}_{-}$. In practices, one can
straightforwardly substitute the expression of density Eq. \ref{density}
into the interaction energy Eq. \ref{int}, and the interaction energy $%
\mathcal{E}_{I}$ becomes a function of $\alpha $, $\mathbf{p}_{\pm }$ and $%
\theta _{\pm }$. Then by minimizing the total energy respect to $%
\alpha $, $\mathbf{p}_{\pm }$ and $\theta _{\pm }$, one can finally obtain
the ground state condensate wave function for given $\Omega $, $\delta $ and
interaction parameters $g_{\uparrow \uparrow }$, $g_{\downarrow \downarrow }$
and $g_{\uparrow \downarrow }$. Thus, a phase diagram can be constructed.
More details have been discussed in Ref. \cite{NIST_Nature,Ho,Stringari1}
and here we will just emphasize some general features.

(1) If the energy minimization gives $\alpha =0$ or $\alpha =\pi /2$,
there is only one momentum component in the condensate wave function, and
the density for both spin component are uniform, as it can be easily seen by
setting $\alpha =0$ or $\pi /2$ in Eq. (\ref{density}). This is
the \textquotedblleft plane wave" condensate. While if the energy minimization
gives $0<\alpha <\pi /2$, both $n_{\uparrow }(\mathbf{r})$ and $%
n_{\downarrow }(\mathbf{r})$ have spatially periodic modulation. This is
named as \textquotedblleft stripe" condensate \cite{Zhai}.

(2) Because of SO coupling, the densities of both components depend on
momentum. If the interaction is $SU(2)$ invariant, i.e. $g_{\uparrow%
\uparrow}=g_{\downarrow\downarrow}=g_{\uparrow\downarrow}=g$, the
interaction energy becomes $\mathcal{E}_{I}=(g/2)\int d^3\mathbf{r}
(n_{\uparrow}(\mathbf{r})+n_{\downarrow}(\mathbf{r}))^2$. In this case, if
the condensate is in the ''plane wave" phase, the interaction energy is
independent of momentum, and one can take $\mathbf{p_{\pm}}$ as $\mathbf{k_{\pm}}$ which
minimizes the single particle energy. However, because the two pseudo-spin
states are taken as two hyperfine level of atoms, they do not have to obey $%
SU(2)$ spin rotational symmetry. Thus, for a general case, three interaction
parameters $g_{\uparrow\uparrow}$, $g_{\downarrow\downarrow}$ and $%
g_{\uparrow\downarrow}$ are all unequal. Therefore, even for the plane wave
condensate, the interaction energy also depends on momentum $p_{\pm}$. That
is equivalent to say, even at mean-field level the self-energy correction
has momentum dependence, which effectively modifies the single particle dispersion and
changes the location of its minimum. In contrast, without SO coupling,
the mean-field self-energy correction is just a constant shift of single
particle energy. Thus, as first pointed out by Ref. \cite{Stringari1},
$\mathbf{p_{\pm}}$ is shifted away from single particle minimum $\mathbf{k_{\pm}}$ \footnote{%
However, for some special case, such as pure Rashba SO coupling without
Zeeman field considered in Ref. \cite{Zhai}, the mean-field self-energy
correlation is also momentum independent.}.

(3) The stripe phase has distinct properties from plane-wave phase. For the stripe phase, the total density
will have spatial modulation, because the spin wave function at $\mathbf{p}_{+}$ and $\mathbf{p}_{-}$ are not orthogonal, i.e., $\cos\theta_\mathbf{p_+}%
\cos\theta_\mathbf{p_-}+\sin\theta_\mathbf{p_+}\sin\theta_\mathbf{p_-}\neq 0$\footnote{%
This is also different from pure Rashba case where the spin wave function
are orthogonal for opposite momentum, and the total density is uniform for
stripe phase}. Thus, in this system the stripe phase is not favored by
density interaction part $(g/2)\int d^3\mathbf{r} (n_{\uparrow}(\mathbf{r}%
)+n_{\downarrow}(\mathbf{r}))^2$. In another word, if the interaction is $%
SU(2)$ invariant, stripe phase will not exist in this system. The difference
in $g_{\uparrow\uparrow}$, $g_{\downarrow\downarrow}$ and $%
g_{\uparrow\downarrow}$ are necessary for stabilizing the stripe phase.
Moreover, since the non-uniform term in total density increases as $\Omega$
increases. Thus, the stripe phase, if exists, should be found in small $%
\Omega$ regime of the phase diagram.

(4) Consider the limit $\Omega=0$, if $g_{\uparrow\uparrow}g_{\downarrow%
\downarrow}-g^2_{\uparrow\downarrow}>0$, a homogeneous mixture of two components is stable against local density fluctuations, and there will be a mixed phase within certain detuning window. Such a mixed phase will turn into stripe phase
once $\Omega$ becomes non-zero, for instance, for $^{87}$Rb case in Fig. 1(a).
While if $g_{\uparrow\uparrow}g_{\downarrow\downarrow}-g^2_{\uparrow\downarrow}<0$,
the mixed phase is not stable against phase separation even for zero $\Omega$, and
there will be no stripe phase in the phase diagram, for instance, for $^{23}$Na case in Fig. 1(b).


Hereafter, we should focus only on $SU(2)$ invariant interaction. This is
relevant for experiment with Rb or Na, because the difference in $%
g_{\uparrow \uparrow }$, $g_{\downarrow \downarrow }$ and $g_{\uparrow
\downarrow }$ are smaller than $1\%$. The generalization to non-$%
SU(2)$ interaction is straightforward. Besides, we only consider the plane
wave phase, because in these systems, the stripe phase either occupies a
very small regime of phase diagram or does not exist.

\begin{figure}[tbp]
\includegraphics[width=5 in]{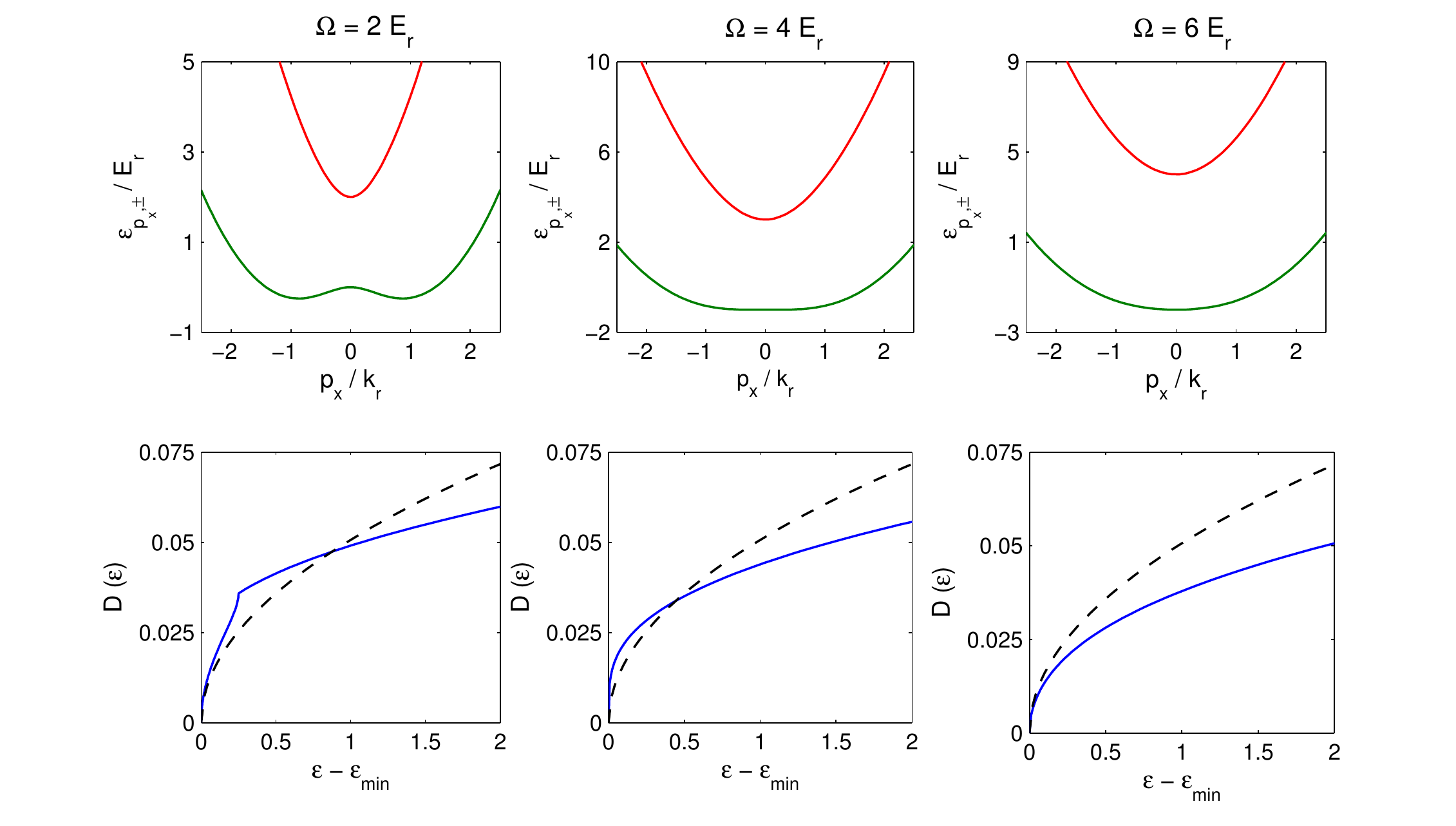}
\caption{Upper panel: Single particle dispersion $\protect\varepsilon%
_{\pm}(p_x) $ as a function of $p_x/k_r$; Lower panel: DoS as a function of $%
\protect\varepsilon-\protect\varepsilon_{\rm min}$ for $\Omega=2E_r$, $4E_r$ and $6E_r$%
, respectively. $\protect\varepsilon_{\rm pmin}$ denotes the minimum value
of single particle energy. The dashed lines are DoS for the case without Raman
coupling.} \label{single_particle}
\end{figure}

\subsection{Zero-detuning case}

In this work we will particularly focus on the case with $\delta=0$ for
following two reasons.

(1) \textit{Density-of-State Effect}. When $\Omega <4E_{r}$, $\epsilon
_{-}(k_{x})$ has two minima at $k_{\pm }=\pm k_{r}\sqrt{1-\left( \frac{%
\Omega }{4E_{r}}\right) ^{2}}$, and when $\Omega >4E_{r}$, $\epsilon
_{-}(k_{x})$ has one single minimum at $k_{x}=0$. Expanding the dispersion
around its minimum, one gets the effective mass in $x$ direction as
\begin{equation}
m^{\ast }=\left\{
\begin{array}{c}
m\left( 1-\frac{\Omega ^{2}}{16E_{r}^{2}}\right)^{-1} \ \ \Omega <4E_{r}
\\
m\left( 1-\frac{4E_{r}}{\Omega }\right)^{-1} \ \ \Omega >4E_{r}%
\end{array}%
\right. .\label{meff}
\end{equation}%
Hence, the low-energy DoS increases with $\Omega $ when $\Omega <4E_{r}$ and
decreases with $\Omega $ when $\Omega >4E_{r}$, as shown in Fig. \ref%
{single_particle}. The most intriguing point is at $\Omega =4E_{r}$ when
the single particle dispersion behaves as $\sim p_{x}^{4}$ at the lowest
order and the low-energy DoS reaches its maximum. As we shall see in later
discussion, this has important physical consequences in the superfluid critical
velocity and the BEC transition temperature.

(2) \textit{$Z_2$ Symmetry and Magnetization}. When $\Omega<4E_r$, bosons
condense into one of the minima, which breaks the $Z_2$ symmetry. The Bose
condensate will have finite magnetization. While when $\Omega>4E_r$, bosons
condense at zero-momentum state and the condensate is non-magnetic. Thus,
there will be a magnetic phase transition at $\Omega=4E_r$ associated with the $%
Z_2$ symmetry breaking, and a divergent spin susceptibility has been predicted
and experimentally found  \cite{USTC_boson, Stringari3}. We note that
such a transition exists only for $\delta=0$, since for non-zero $\delta$
the Hamiltonian does not possess the $Z_2$ symmetry, and the condensate
phase is always magnetic.

\section{Bogoliubov Theory and Superfluid Critical Velocity}

\subsection{Bogoliubov Spectrum}

We study the fluctuations around the condensate with Bogoliubov theory.
Considering a plane-wave condensate at  momentum $\mathbf{p%
}_{0}=p_{0}\mathbf{e}_{x}$, the field operator can be expanded as
\begin{equation}
\hat{\psi}(\mathbf{r})=\varphi(\mathbf{r}) +\delta \hat{\psi}(\mathbf{r})
\end{equation}%
where $\varphi(\mathbf{r})$ is the condensate wave-function
\begin{equation}
\varphi(\mathbf{r}) =\sqrt{n_{0}}\left(
\begin{array}{c}
\cos \theta_{p_0} \\
-\sin \theta_{p_0} %
\end{array}%
\right) \exp \left( ip_{0}x\right)
\end{equation}%
and satisfies the Gross-Pitaevskii (GP) equation
\begin{equation}
\Big[H_{0}(p_{0})+gn_0\mathbf{I}\Big]\left(
\begin{array}{c}
\cos \theta_{p_0}  \\
-\sin \theta_{p_0} %
\end{array}%
\right) =\mu \left(
\begin{array}{c}
\cos \theta_{p_0}  \\
-\sin \theta_{p_0} %
\end{array}%
\right).
\end{equation}%

Defining $\hat{\Psi}_{\mathbf{q}}^{\dag }=(\psi _{\mathbf{p}_{0}+\mathbf{q}%
,\uparrow }^{\dag },\psi _{\mathbf{p}_{0}+\mathbf{q},\downarrow }^{\dag
},\psi _{\mathbf{p}_{0}-\mathbf{q},\uparrow },\psi _{\mathbf{p}_{0}-\mathbf{q%
},\downarrow })$, the Bogoliubov Hamiltonian for the fluctuation part can be
written as
\[
\mathcal{K}=\sum_{q_{x}>0}\hat{\Psi}_{\mathbf{q}}^{\dag }\mathcal{K}_{%
\mathbf{q}}\hat{\Psi}_{\mathbf{q}}-\frac{1}{2}\sum_{q_{x}>0}\big[\epsilon _{%
\mathbf{p_{0}-q,\uparrow }}+\epsilon _{\mathbf{p_{0}-q,\downarrow }}-2\mu
+3gn_{0}\big]
\]%
where
\begin{equation}
\mathcal{K}_{\mathbf{q}}=\left(
\begin{array}{cc}
K_{0}(\mathbf{p}_{0}+\mathbf{q})+\Sigma _{N} & \Sigma _{A} \\
\Sigma _{A} & K_{0}(\mathbf{p}_{0}-\mathbf{q})+\Sigma _{N}%
\end{array}%
\right)
\end{equation}%
$K_{0}=H_{0}-\mu $ is the grand Hamiltonian of non-interacting system, $\Sigma _{%
\mathrm{N}}$ and $\Sigma _{\mathrm{A}}$ are normal and anomalous self-energy
respectively
\begin{equation}
\Sigma _{\mathrm{N}}=g n_0 \left(
\begin{array}{cc}
\sin ^{2}\theta_{p_0} +2\cos ^{2}\theta_{p_0}  & -\sin \theta_{p_0} \cos \theta_{p_0} \\
-\sin \theta_{p_0} \cos \theta_{p_0} & \cos ^{2}\theta_{p_0} +2\sin ^{2}\theta_{p_0} %
\end{array}%
\right) ;
\end{equation}%
\begin{equation}
\Sigma _{\mathrm{A}}=g n_{0}\left(
\begin{array}{cc}
\cos ^{2}\theta_{p_0} & -\sin \theta_{p_0} \cos \theta_{p_0} \\
-\sin \theta_{p_0} \cos \theta_{p_0} & \sin ^{2}\theta_{p_0}%
\end{array}%
\right) .
\end{equation}%
Hence the Bogoliubov spectrum is determined by
\begin{equation}
\mathrm{Det}\left(
\begin{array}{cc}
K_{0}(\mathbf{p}_{0}+\mathbf{q})+\Sigma _{\mathrm{N}}-E & \Sigma _{\mathrm{A}%
} \\
-\Sigma _{\mathrm{A}} & -K_{0}(\mathbf{p}_{0}-\mathbf{q})-\Sigma _{\mathrm{N}%
}-E%
\end{array}%
\right) =0  \label{Det_eq}
\end{equation}

Because of the emergence of off-diagonal long range order, the excitation
energy should be gapless in the long wave-length limit $\mathbf{q}%
\rightarrow 0$, which requires
\begin{equation}
\mathrm{Det}\Big[\Sigma_{\mathrm{A}}\big(K_{0}(p_0)+\Sigma_{\mathrm{N}}\big)%
^{-1}\Sigma_{\mathrm{A}}-\big(K_{0}(p_0)+\Sigma_{\mathrm{N}}\big) \Big]=0
\end{equation}
This equation is indeed satisfied because GP equation for the condensate
wave-function can be rewritten as
\begin{equation}
\mathrm{Det} \Big[ H_{0}(p_0)-\mu+\Sigma_{\mathrm{N}}-\Sigma_{\mathrm{A}} %
\Big] =0  \label{HP_relation}
\end{equation}
Eq. (\ref{HP_relation}) can also be regarded as a generalization of
Hugenholtz-Pines relation $\mu=\Sigma_{\mathrm{N}}-\Sigma_{\mathrm{A}}$ to
the SO coupled Bose gases \cite{Hugenholtz-Pines}.

\begin{figure}[tbp]
\includegraphics[width=5 in]{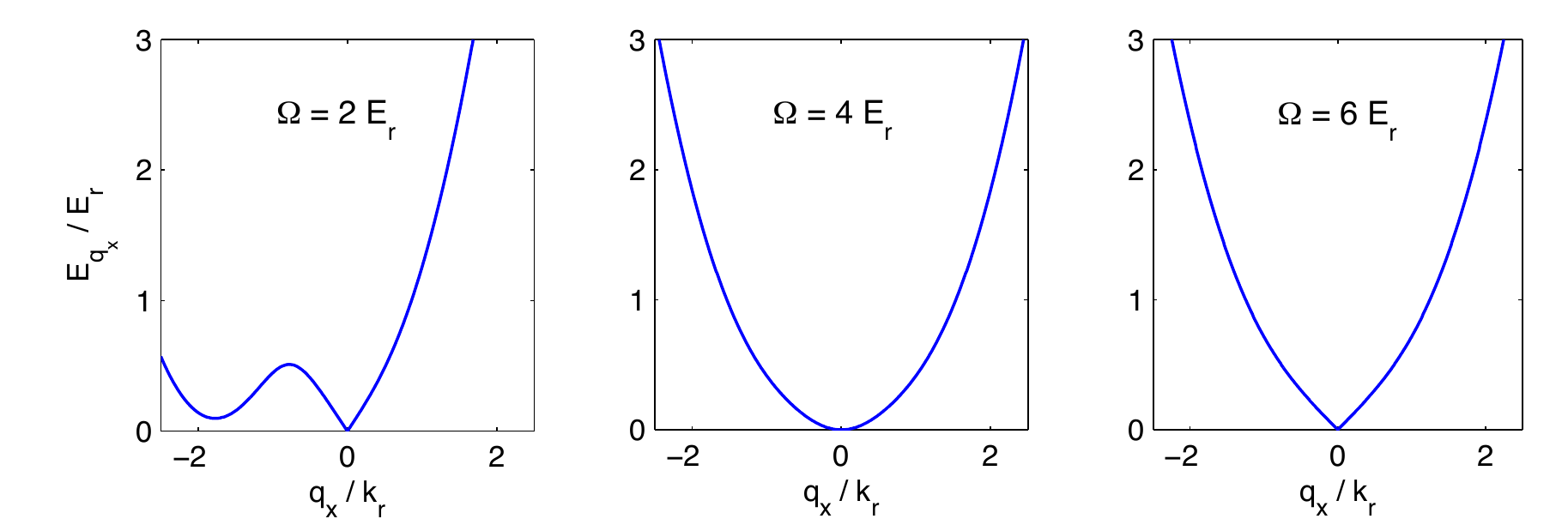}
\caption{Bogoliubov spectrum for $\Omega/E_{\mathrm{r}}=2,4$ and $6$ for
(a-c). $gn/E_{\mathrm{r}}$ is fixed to be $0.5$. }
\label{Bogoliubov}
\end{figure}

Solving Eq. (\ref{Det_eq}) gives rise to the entire Bogoliubov spectrum $E_{%
\mathbf{q},\pm }$. Examples are displayed in Fig. \ref{Bogoliubov} for
various $\Omega /E_{\mathrm{r}}$. For $\mathbf{q}_x\rightarrow 0$, the
low-energy excitation is the phonon mode with a linear dispersion $E_{-}(q_x)=s q_x$. Along the direction of Raman beam, the sound velocity is given by
\begin{equation}
s=\sqrt{\frac{gn_{0}}{m}\Big(1-{\frac{\Omega ^{2}}{16E_{\mathrm{r}}^{2}}}%
\Big)} \label{s1}
\end{equation}%
when $\Omega <4E_{\mathrm{r}}$ and
\begin{equation}
s=\sqrt{\frac{gn_{0}}{m}\Big(1-{\frac{4E_{\mathrm{r}}}{\Omega
}}\Big)} \label{s1}
\end{equation}%
when $\Omega >4E_{\mathrm{r}}$. These results are consistent with the effective
mass approximation that gives $s=\sqrt{gn_0/m^*}$. At $\Omega =4E_{\mathrm{r}}$, the effective mass
diverges, and the low-energy phonon mode is quadratic in $q_{x}^{2}$, as
shown in Fig. \ref{Bogoliubov}(b). This is a significant difference compared
to the conventional phonon mode, which is always linear in $\mathbf{q}$.
When $\Omega <4E_{\mathrm{r}}$, the Bogoliubov spectrum has another local
minimum at finite $q_{x}$, which is due to the double degeneracy in single
particle spectrum. This is attributed as \textquotedblleft roton minimum" by
Ref. \cite{Stringari3}.

\subsection{Quantum and Thermal Condensate Depletion}

The Bogoliubov Hamiltonian can be straightforwardly written as
\begin{eqnarray}
&&\mathcal{K}^{(2)}=\sum_{\mathbf{q}\neq 0}\big(E_{\mathbf{q,+}}\hat{\tilde{%
\psi}}_{\mathbf{q,+}}^{\dag }\hat{\tilde{\psi}}_{\mathbf{q,+}}+E_{\mathbf{q,-%
}}\hat{\tilde{\psi}}_{\mathbf{q,-}}^{\dag }\hat{\tilde{\psi}}_{\mathbf{q,-}}%
\big)  \nonumber \\
&&+{\frac{1}{2}}\sum_{\mathbf{q}\neq 0}\Big[E_{-\mathbf{q,+}}+E_{-\mathbf{q,-%
}}-\epsilon _{\mathbf{p_{0}-q,\uparrow }}-\epsilon _{\mathbf{%
p_{0}-q,\downarrow }}+2\mu -3gn_{0}\Big]  \label{diagonalized_K}
\end{eqnarray}%
here $\hat{\tilde{\Psi}}_{\mathbf{q}}^{\dag }\equiv (\hat{\tilde{\psi}}_{%
\mathbf{q,+}}^{\dag },\hat{\tilde{\psi}}_{\mathbf{q,-}}^{\dag },\hat{\tilde{%
\psi}}_{-\mathbf{q,+}},\hat{\tilde{\psi}}_{-\mathbf{q,-}})=M_{\bf q}^{-1} \hat{\tilde{\Psi}}_{\mathbf{q}}$ are quasi-particle operators, and $M_{\mathbf{q}}$ is the Bogoliubov transformation matrix which is obtained from eigen equation (\ref{Det_eq}). The quantum
and thermal depletion of condensate $n_{\rm ex,\sigma}={1\over V}\sum_{{\bf q }\neq 0}\langle\hat\psi_{{\bf p}_0+{\bf q},\sigma}^\dag\hat\psi_{{\bf p}_0+{\bf q},\sigma}\rangle$ thus can be readily calculated using $M_{\mathbf{q}}$ and $E_{\mathbf{%
q},\pm }$.

\begin{figure}[tbp]
\includegraphics[width=5 in]{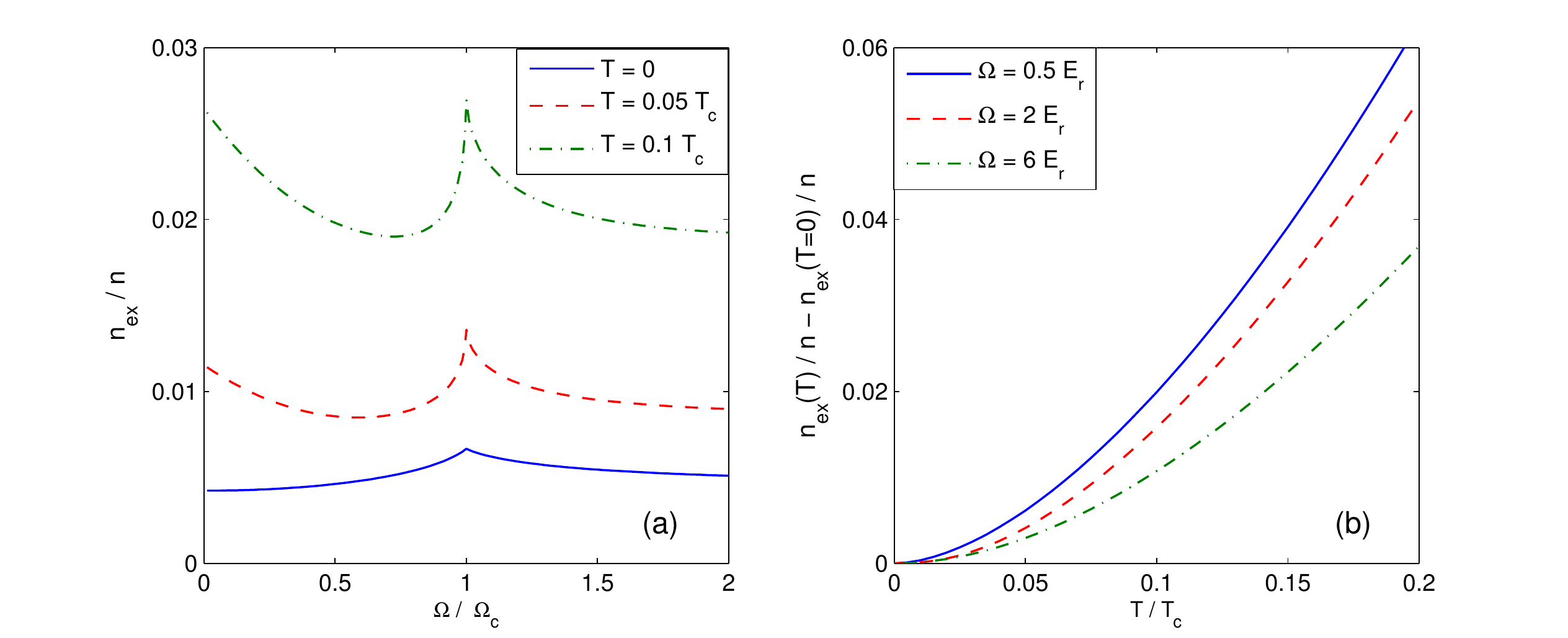}
\caption{ (a) Total depletion as a function of $\Omega/E_r$
for $T=0, 0.05T_{\mathrm{c}}$ and $0.1 T_{\mathrm{c}}$, where $T_c$ is the condensation temperature (see Section 4). (b)
Thermal depletion as a function of $T/T_{\mathrm{c}}$ for various values of $%
\Omega/E_r$ (take $gn/E_r=0.5$ and $n/k_r^3=1$). }
\label{Depeltion}
\end{figure}

The depletion fraction as a function of Raman coupling strength $\Omega$ and temperature $T$ is shown in Fig. \ref{Depeltion}. The contribution to
depletion is mainly from two parts: the low-energy phonon part and the roton
part. At zero-temperature, the contribution is dominated by the phonon part
because the roton part has a finite excitation gap. Thus, due to the non-monotonic
behavior of sound velocity discussed above, when $\Omega <4E_r$, the quantum depletion $n_{\mathrm{ex}}/n$ increases with
$\Omega $,
 reaches a maximum at $\Omega =4E_r$, and then decreases as $\Omega $ increases. At finite temperature,
because the roton gap is quite small at small $\Omega $, the roton part will
give an significant contribution. For $\Omega <4E_r$, the
contribution from phonon part always increases with $\Omega $ because the
decrease of phonon velocity, while the contribution from roton part
decreases with $\Omega $ because the roton gap increases with $\Omega $. Due
to the interplay of these two contributions, for small $\Omega $, $n_{%
\mathrm{ex}}/n$ first decreases as $\Omega $ increase, in contrast
to
zero-temperature case. Then $n_{\mathrm{ex}}/n$ increases again with $%
\Omega $ and the peak of $n_{\mathrm{ex}}/n$ at $\Omega =4E_r$ retains.

\begin{figure}[tbp]
\includegraphics[width=5 in]{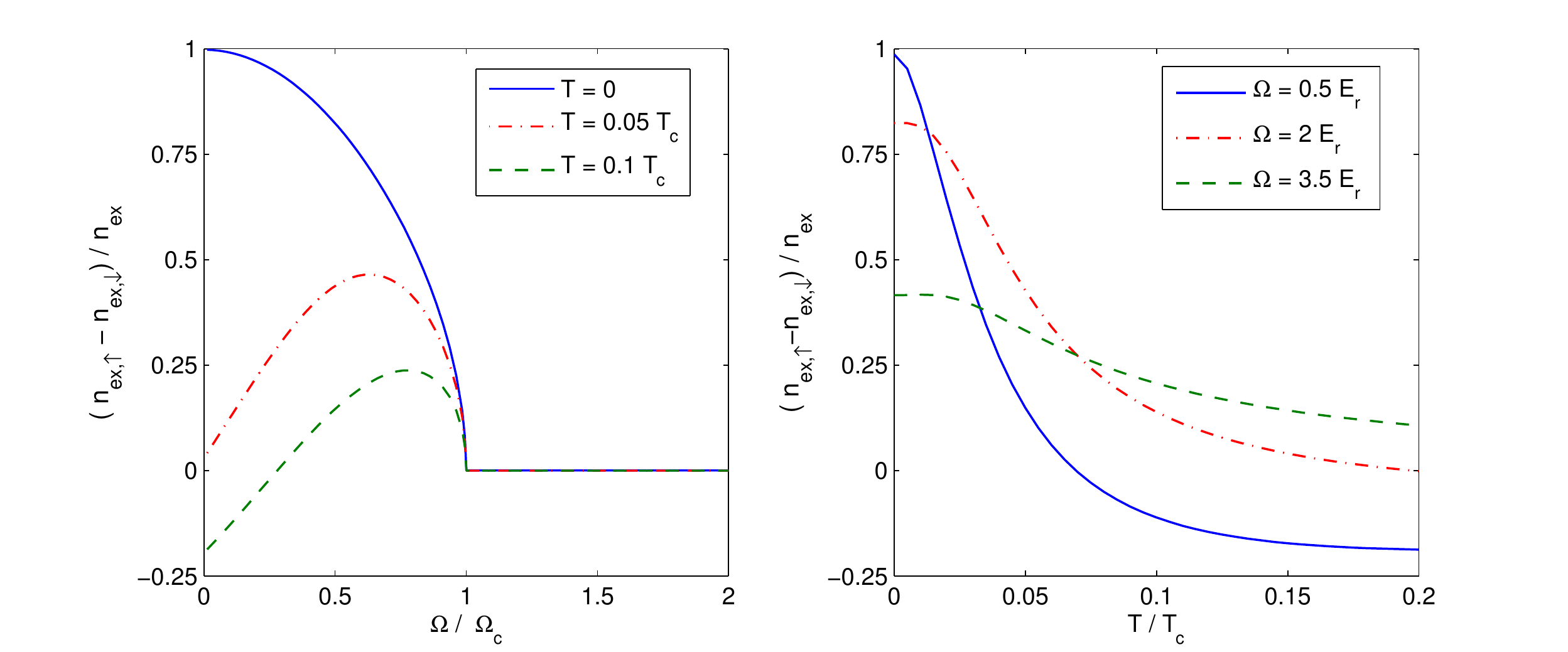}
\caption{ (a) Magnetization of depletion $(n_{\mathrm{ex}\uparrow}-n_{\mathrm{ex}\downarrow})/n_{%
\mathrm{ex}}$ as a function of $\Omega/E_r$ for
various
temperatures. (b) $(n_{\mathrm{ex}\uparrow}-n_{\mathrm{ex}\downarrow})/n_{%
\mathrm{ex}}$ as a function of $T/T_{\mathrm{c}}$ for various
$\Omega$. (parameters are the same as in Fig.
\protect\ref{Depeltion}) } \label{magnetization}
\end{figure}

Another interesting manifestation of roton minimum is through the
magnetization of non-condensed part defined as
$(n_{\mathrm{ex}\uparrow }-n_{\mathrm{ex}\downarrow
})/n_{\mathrm{ex}}$. Assuming the condensate
momentum $p_{0}>0$, the condensate has a positive magnetization $%
(n_{0\uparrow }-n_{0\downarrow })/n_{0}>0$. As shown in Fig. \ref%
{magnetization} (a), at zero-temperature, the depletion also has a positive
magnetization $(n_{\mathrm{ex}\uparrow }-n_{\mathrm{ex}\downarrow })/n_{%
\mathrm{ex}}>0$. This is because the phonon contribution, which
dominates the quantum depletion, originates from states whose
momenta are nearby condensate momentum and possess same
magnetization as the condensate. However, at finite temperature, for small $%
\Omega $, the depletion can have opposite magnetizations as the condensate part,
as shown in Fig. \ref{magnetization}(a). And from Fig. \ref{magnetization}(b)
one can also see that, for instance at $\Omega =0.5E_r$,
the magnetization soon decreases to negative as temperature increases. This
is because the roton contribution dominates in this regime. Recall that when
$\Omega <4E_r$ there is double minimum in this single
particle spectrum with opposite momentum and magnetization. When condensate
takes place in one of the minimum, the roton minimum in the excitation
appears nearby the other minimum. Thus, the contribution from roton minimum
displays opposite magnetization with condensate part.
As we know, the phonon dispersion is linear, while the roton dispersion is quadratic.
That implies the DoS of the quasi-particle near the roton is larger than the one near the phonon part.
At finite temperatures higher than the roton gap, the thermal depletion near the roton minimum will
exceed the depletion from phonon modes. This leads to an opposite magnetization of the
total thermal depletion compared with the condensate. In
the further experiment, measuring the magnetization of the thermal depletion
will thus provide an evidence for the exsistence of the roton minimum in
the excitation spectrum.

\subsection{Beyond-mean-field correction of ground state energy}

Bogoliubov theory also gives rise to a correction to mean-field energy
usually named as Lee-Huang-Yang correction \cite{LHY}. The Lee-Huang-Yang
correction can be considered as a sum of all zero-point energies of excitation
modes at different momenta. Without SO coupling, it is a function of $%
n^{1/3}a_{\mathrm{s}}$ only and the ratio between energy correction to mean-field energy is
\begin{equation}
\frac{\mathcal{E}_{\mathrm{LHY}}}{\mathcal{E}_{\mathrm{MF}}}=\Gamma _{%
\mathrm{LHY}}(n^{1/3}a_{\mathrm{s}})={\frac{128}{15\sqrt{\pi }}}\sqrt{%
na_{s}^{3}}.
\end{equation}%
\begin{figure}[tbp]
\includegraphics[width=5 in]{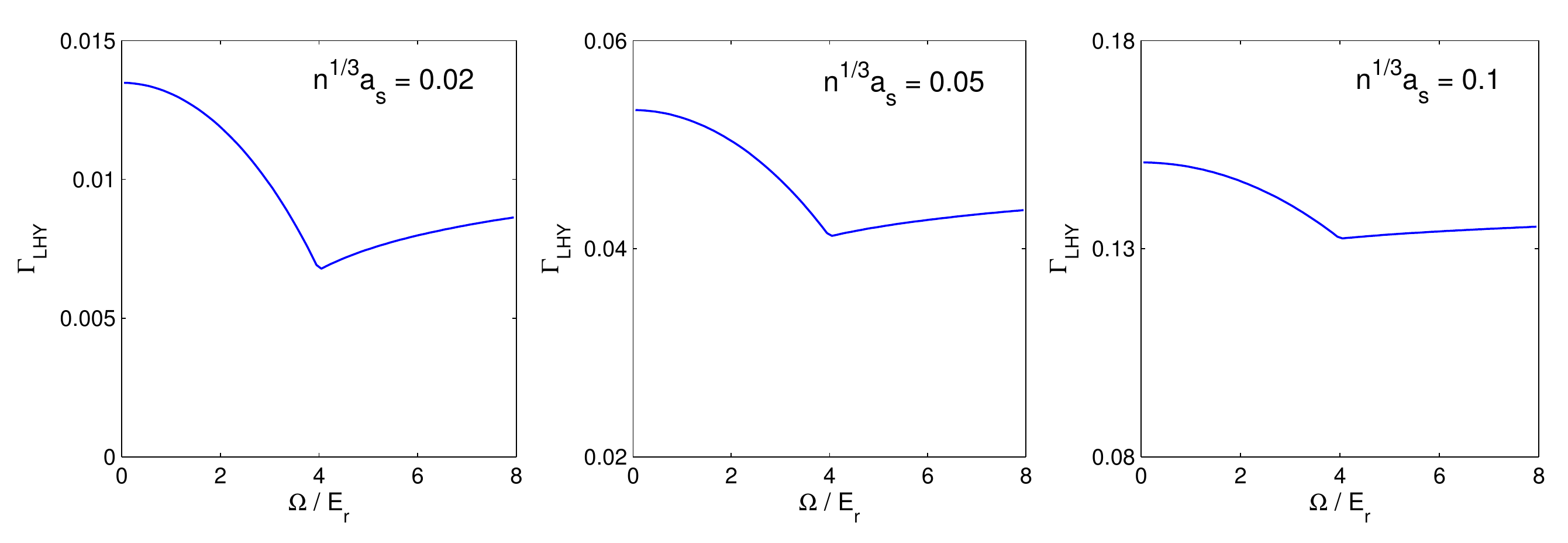}
\caption{$\Gamma_{\mathrm{LHY}}$ as a function of $\Omega/E_{r}$
for different $n^{1/3}a_{\mathrm{s}}$ (fixed $n/k_{\mathrm{r}}^3=1$). }
\label{LHY}
\end{figure}
While with SO coupling, using Eq. (\ref{diagonalized_K}) one finds that $%
\mathcal{E}_{\mathrm{LHY}}$ is given by
\begin{eqnarray}
{\mathcal{E}_{\mathrm{LHY}}} &=&{\frac{1}{2}}\sum_{\mathbf{q}\neq 0}\Big[E_{-%
\mathbf{q,+}}+E_{-\mathbf{q,-}}-\epsilon _{\mathbf{p_{0}-q,\uparrow }%
}-\epsilon _{\mathbf{p_{0}-q,\downarrow }}+2\mu -{\frac{12\pi a_{s}n}{m}}%
\Big]  \nonumber \\
&&+{\frac{8\pi ^{2}a_{s}^{2}n^{2}}{m}}\sum_{\mathbf{k}}{\frac{1}{k^{2}}}
\end{eqnarray}%
where the last term is obtained from the renormalization relation of
coupling constant $g$ via second-order Bohn approximation
\begin{equation}
g={\frac{4\pi a_{s}}{m}}+{\frac{(4\pi a_{s})^{2}}{mV}}\sum_{\mathbf{k}}{%
\frac{1}{k^{2}}}
\end{equation}%
Therefore $\Gamma _{\mathrm{LHY}}$ is a function of $n^{1/3}a_{\mathrm{s}}$,
$\Omega /E_{\mathrm{r}}$ and $n/k_{\mathrm{r}}^{3}$, and the results are
shown in Fig. \ref{LHY}. We find that the Lee-Huang-Yang correction shows a nonmontonic behavior as a function of Raman coupling strength and  displays
a minimum at $\Omega _{\mathrm{c}}=4E_{\mathrm{r}}$. This is consistent with
the softening of phonon mode at $\Omega _{\mathrm{c}}$ discussed above.
The softer the phonon mode is, the smaller the zero-point energy will be, which
results in the minimum in Lee-Huang-Yang correction.

\subsection{Superfluid Critical Velocity}

There are two ways to measure critical velocity \cite{Biao_Wu}. First, the
condensate is at rest and an impurity moves with finite velocity in the
condensate. In this way, one can obtain a critical dragging velocity

\begin{equation}
\mathbf{v}_{\mathrm{drag}}=\min \left( \frac{E_{\mathbf{q}}}{\mathbf{q}}%
\right) ,  \label{v_drag}
\end{equation}%
where $E_{\mathbf{q}}$ is the quasi-particle energy for momentum $\mathbf{q}$.
 Critical dragging velocity is shown in Fig. 6(a). One can see that $\mathbf{v%
}_{\mathrm{drag}}$ vanishes at $4E_{r}$, because the phonon
spectrum becomes quadratic in $q_{x}$ at this point and the phonon velocity
vanishes. One can also see that for $\Omega <4E_{r}$, $\mathbf{%
v}_{\mathrm{drag}}$ is not symmetric for moving along $\hat{x}$ or along $-%
\hat{x}$. This is due to the non-symmetric structure of Bogoliubov spectrum
as shown in Fig. \ref{Bogoliubov}(a). Moving along $\hat{x}$, $\mathbf{v}_{%
\mathrm{drag}}$ is determined by phonon part. Moving along $-\hat{x}$, $%
\mathbf{v}_{\mathrm{drag}}$ is determined by the the roton minimum for small $\Omega$. Since
the roton gap increases with $\Omega $, the $\mathbf{v}_{\mathrm{drag}%
}$ along $-\hat{x}$ also increase with $\Omega $ untill the roton vanishes. After
that the $\mathbf{v}_{\mathrm{drag}}$ along $-\hat{x}$ is also determined by
the phonon mode, and will decrease with $\Omega $ until $\Omega =4E_{r}$.

\begin{figure}[tbp]
\includegraphics[width=5 in]{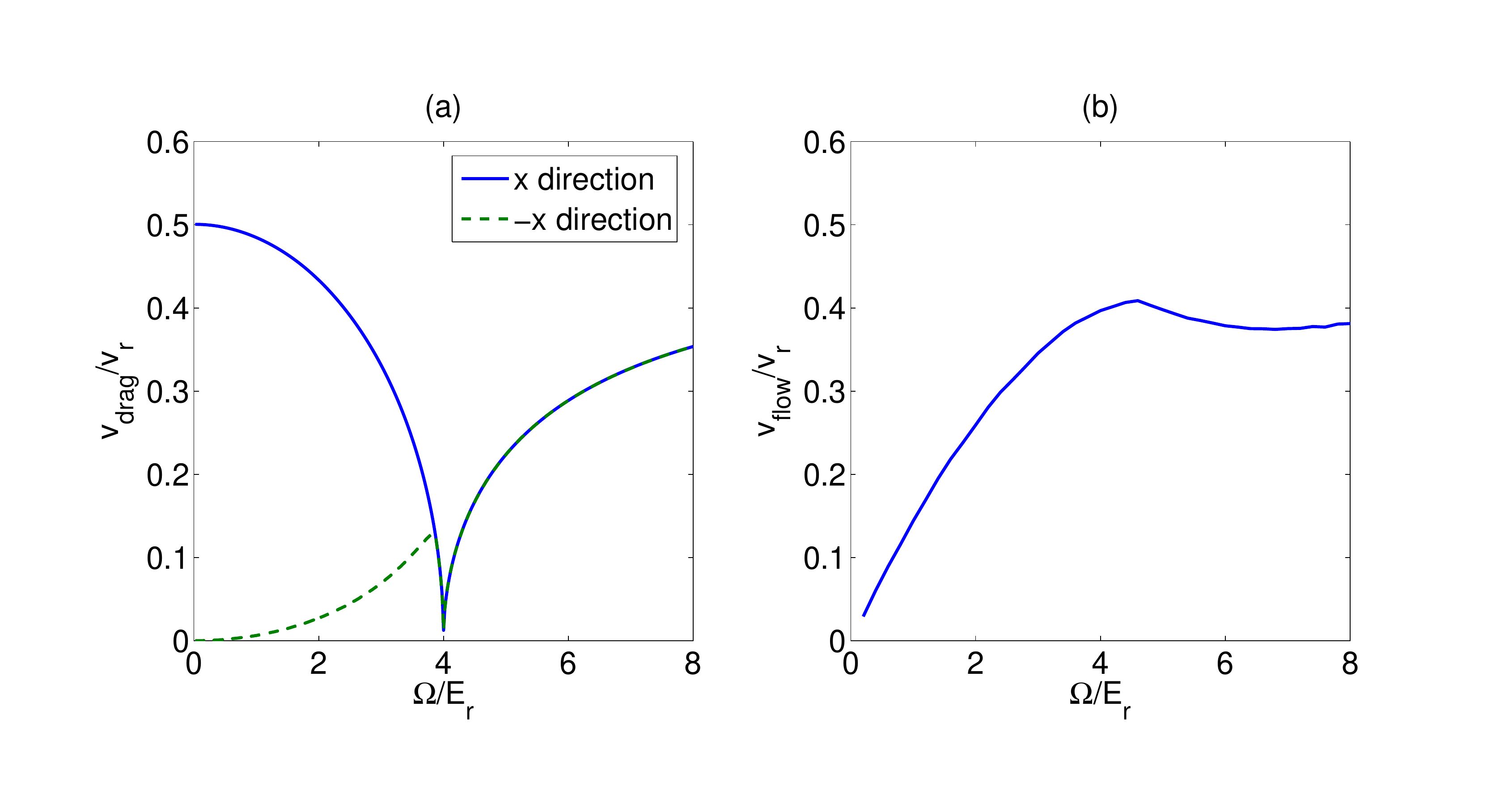}
\caption{ Critical dragging(a) and flowing(b) velocity as a function of $\Omega/E_{\mathrm{r}}$.
Here $gn=0.5E_{r}$, and $v_r=k_r/m$. }
\label{critical_velocity}
\end{figure}

Second, one can consider a static impurity and let the condensate move with
a finite velocity. This determines another critical velocity which can be
called critical flowing velocity $\mathbf{v}_{\mathrm{flow}}$. If the system
is Galilean invariant, $\mathbf{v}_{\mathrm{drag}}$ and $\mathbf{v}_{\mathrm{flow}}$ should be identical. However, SO
coupling breaks the Galilean invariance, and thus these two velocities
become unequal, as first pointed out in Ref. \cite{Biao_Wu} for Rashba type
SO coupling. In our system, Galilean invariance is also broken. Considering
a Galilean transformation in $\hat{x}$ direction, the Hamiltonian in the
moving frame becomes
\begin{eqnarray}
H_{0}\left( \mathbf{v}\right) &=&\frac{1}{2m}\left\{ \left( \hat{p}%
_{x}-k_{r}\sigma _{z}\right) ^{2}\right\} +\frac{1}{2}\Omega \sigma
_{x}-v_{x}p_{x}  \label{Gal_TF} \\
&=&\frac{1}{2m}\left[ \hat{p}_{x}-\left( k_{r}\sigma _{z}+mv_{x}\right) %
\right] ^{2}+\frac{1}{2}\Omega \sigma _{x}-v_{x}k_{r}\sigma _{z}-\frac{1}{2}%
mv^{2}  \label{mov_H}
\end{eqnarray}%
With a gauge transformation $U\left( \mathbf{x},t\right) =\exp \left[
-imv_{x}x-i\left( mv^{2}/2\right) t\right] $, the Hamiltonian becomes
\begin{equation}
H_{0}\left( \mathbf{v}\right) =\frac{1}{2m}\left( \hat{p}_{x}-k_{r}\sigma
_{z}\right) ^{2}+\frac{1}{2}\Omega \sigma _{x}-v_{x}k_{r}\sigma _{z}
\end{equation}%
Compared to the Hamiltonian in the stationary frame, there is an additional
velocity dependent Zeeman term $v_{x}k_{r}\sigma _{z}$. This term can not be
gauged away, and that implies the broken of Galilean invariance for a SO
coupled particle. The physical effect of this term has already been observed
in Ref. \cite{USTC_boson} in collective dipole oscillation experiment. $%
\mathbf{v}_{\mathrm{flow}}\neq \mathbf{v}_{\mathrm{drag}}$ is another
manifestion of the absence of Galilean invariance.

To determine $\mathbf{v}_{\mathrm{flow}}$ we shall first find out the ground
state wave function for the Hamiltonian in the comoving frame [Eq. (\ref{mov_H})], say
\begin{equation}
\varphi ^{\prime }(\mathbf{x})=\sqrt{n_{0}}\left(
\begin{array}{c}
\cos \theta ^{\prime } \\
-\sin \theta ^{\prime }%
\end{array}%
\right) \exp \left( i\mathbf{p}_{0}^{\prime }\mathbf{x}\right) .
\label{mov_WF}
\end{equation}%
Then, following similar procedure discussed above, one can find out the
Bogoliubov spectrum above this new ground state in the comoving frame, denoted
by $E_{\mathbf{q}}^{\prime }\left( \mathbf{v}\right) $. Then, observing in
the laboratory frame, the excitation spectrum is given by $E_{\mathbf{q}%
}\left( \mathbf{v}\right) =E_{\mathbf{q}}^{\prime }\left( \mathbf{v}\right) +%
{\mathbf{v}}\cdot {\mathbf{q}}$. When $\mathbf{v}$ is above certain critical
value, $E_{\mathbf{q}}$ will start to have negative part, which indicates
instability of the condensate. This critical values defines $\mathbf{v}_{%
\mathrm{flow}}$. $\mathbf{v}_{\mathrm{flow}}$ of this system is shown in
Fig. 7(b). We find that, strongly in contrast to $\mathbf{v}_{\mathrm{drag}}$%
, $\mathbf{v}_{\mathrm{flow}}$ remains finite across $\Omega =4E_{r}$.
This is because in the comoving frame with finite velocity, the single
particle spectrum is always quadratic around its minimum due to the velocity
dependent Zeeman term, which prevents the softening of phonon mode.

\section{Hatree-Fock theory of normal state and BEC transition temperature}

\subsection{Transition temperature of a noninteracting gas}

\begin{figure}[tbp]
\includegraphics[width=5 in]{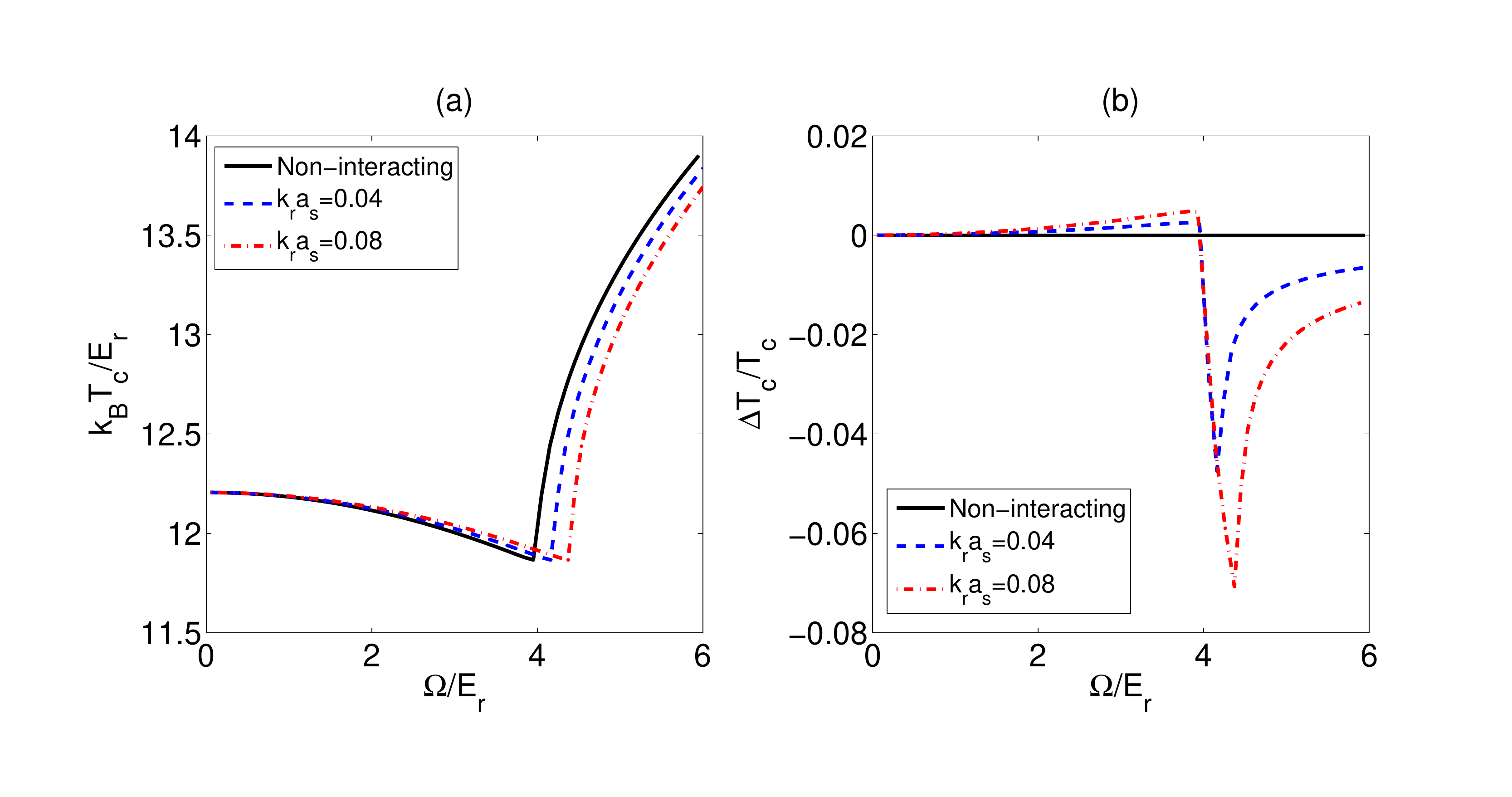}
\caption{ $T_c$ (a) and its relative shift from non-interacting
Bose gases $\Delta T_{\mathrm{c}}/T_{\mathrm{c}}$ (b) of a uniform
system as a function of $\Omega/E_{\mathrm{r}}$ for various
interaction strengths. Here $n/k^3_{\mathrm{r}}=5$. }
\label{Tc_uniform}
\end{figure}

Now we study the normal state of the system and determine the transition
temperature of condensate. First, we consider a noninteracting gas. The
condensation transition temperature $T_{\mathrm{c}}$ of a uniform system is
given by (set $k_{\mathrm{B}}=1$)
\begin{equation}
n=\int_{-\infty }^{\infty }d\varepsilon \frac{D\left( \varepsilon \right) }{%
e^{(\varepsilon -\mu )/T_{c}}-1},
\end{equation}%
where chemical potential reaches the bottom of the single particle spectrum,
$\mu =\epsilon _{\mathrm{min}}$. Due to the DoS effect discussed in Section
1, $T_{\mathrm{c}}$ decreases with $\Omega $ in the regime $\Omega <4E_{r}$,
reaches a minimum of finite value around $\Omega =4E_{r}$, and then increase
in the regime $\Omega >4E_{r}$. This non-monotonic behavior is shown in Fig. %
\ref{Tc_uniform}(a). In $\Omega \rightarrow 0$ limit and $\Omega \rightarrow
\infty $ limit, one can find a simple relation of transition temperature,
\begin{equation}
\frac{T_{c}\left( \Omega \rightarrow \infty \right) }{T_{c}\left( \Omega
\rightarrow 0\right) }=2^{2/3}
\end{equation}%
because the low-energy DoS for the later case shrinks to only half of the
first one.

\begin{figure}[tbp]
\includegraphics[width=5 in]{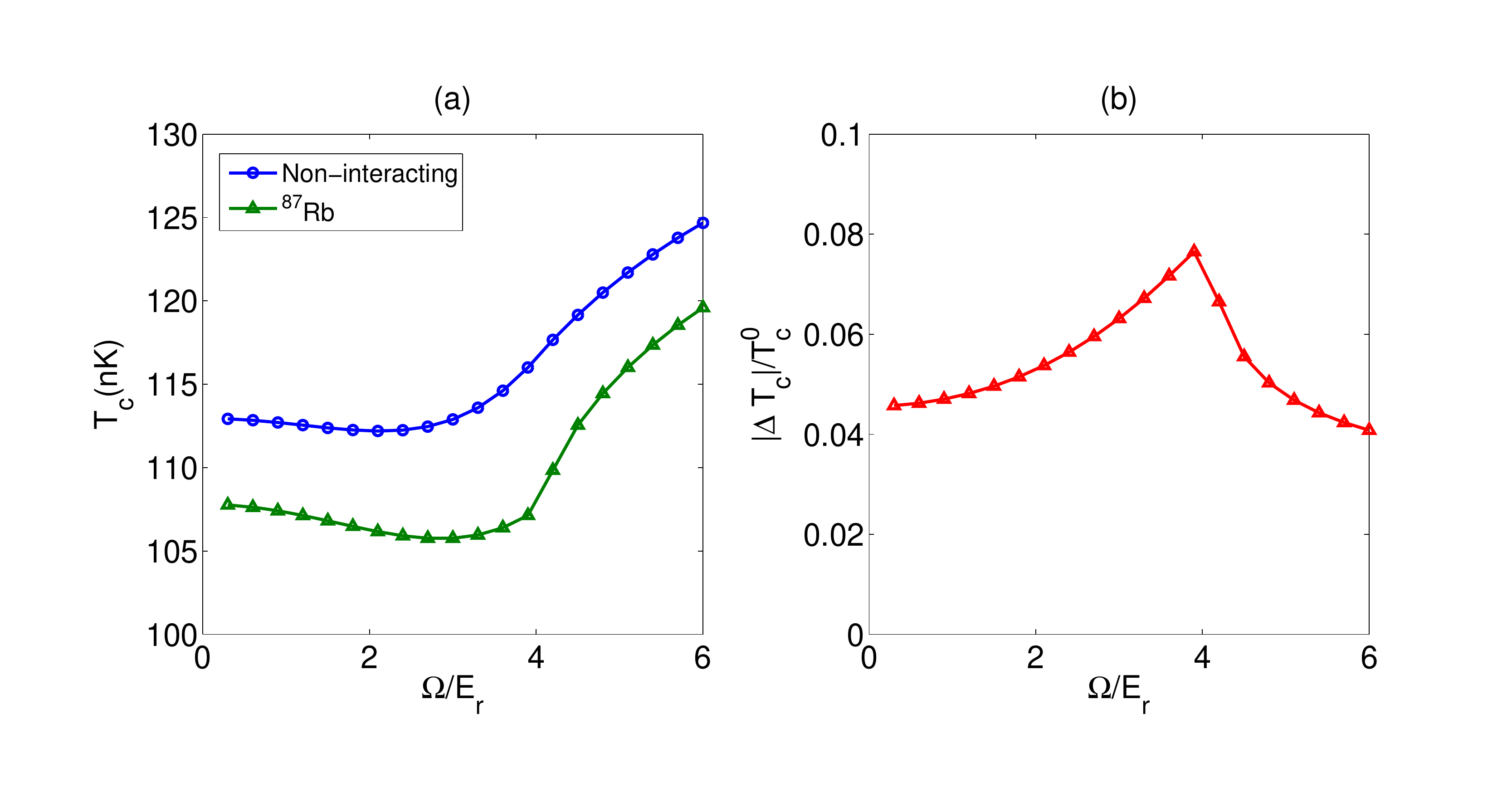}
\caption{(a) $T_c$ of $^{87}$Rb and non-interacting Bose gases in the harmonic trap.
(b) Shift of $T_c$ of $^{87}$Rb due to interaction as a function of $\Omega/E_{r}$.
Here $E_{r}=2\protect\pi \times 2.2\mathrm{kHz}$.
The particle number is $N=2.5\times10^{5}$, and the trapping frequency is $\omega=2\protect\pi \times 50\mathrm{Hz}$.}
\label{Tc_trap}
\end{figure}

In a harmonic trap, the semiclassical energy of single boson can be
expressed as
\begin{equation}
\varepsilon _{\mathbf{p,\pm }}\left( \mathbf{r}\right) =\frac{1}{2m}\left(
\mathbf{p}^{2}+k_{r}^{2}\right) \pm \sqrt{\left( k_{r}p_{x}/m\right)
^{2}+\Omega ^{2}/4}+\frac{1}{2}m\omega ^{2}r^{2},  \label{E_sp_trap}
\end{equation}%
with $\omega $ the trap frequency. With semiclassical approximation, the DoS
should be modified as
\begin{equation}
D_{\mathrm{trap}}(\varepsilon )={\frac{1}{V}}\sum_{\mathbf{p}}\int d^{3}r%
\Big[\delta _{\mathrm{D}}\big(\varepsilon -\varepsilon _{\mathbf{p,+}}(%
\mathbf{r})\big)+\delta _{\mathrm{D}}\big(\varepsilon -\varepsilon _{\mathbf{%
p,-}}(\mathbf{r})\big)\Big]
\end{equation}%
and the transition temperature can be obtained from
\begin{equation}
N=\int_{-\infty }^{\infty }d\varepsilon \frac{D_{\mathrm{trap}}\left(
\varepsilon \right) }{e^{(\varepsilon -\mu )/T_{c}}-1}
\end{equation}%
when chemical potential $\mu $ reaches the minimum of single-particle energy
at trap center. The result is shown in Fig. \ref{Tc_trap}(a).

In contrast to uniform case, one finds the minimum location of the $T_{%
\mathrm{c}}$ is shifted into smaller $\Omega$ regime even for a small particle number. While if the
particle number is large enough, $T_{\mathrm{c}}$ will always monotonically
increases with $\Omega $. This is because the effective mass increases with $%
\Omega $, in a semiclassical sense it will leads an increasing of centre
density in the trap of the thermal gas\cite{Pitaevskii}, $n_{T}\left(
0\right) \propto \left( m^{\ast }T\right) ^{3/2}$. From a quantum view of
point, the effective harmonic length $a_{\mathrm{ho}}^{2}=\frac{\hbar }{%
\sqrt{m^{\ast }m}\omega }$ of the SO coupled Boson decrease with $\Omega $.
That also indicates the increasing of the central density. The density effect
competes with DoS effects, and for large $N$ the former dominates over the latter.
In $\Omega \rightarrow
\infty $ limit, due to the fact that DoS shrinks to one half of that in $\Omega \rightarrow 0$ limit, the transition temperature in these two regimes follows,
\begin{equation}
\frac{T_{c}\left( \Omega \rightarrow \infty \right) }{T_{c}\left( \Omega
\rightarrow 0\right) }=2^{1/3}.
\end{equation}%

\subsection{Mean-field Shift of Transition Temperature for Uniform Gases}

In the absence of SO coupling, it is well known that the contact
interaction between the particles does not affect $T_{\mathrm{c}}$
at mean-field level \cite{Fetter}, this is because the Hatree-Fock
self-energy only provides a constant shift of chemical potential.
While with SO coupling, as shown in the following, the interactions
do have a non-trivial effect even at mean-field level.

To construct a self-consistent Hartree-Fock theory, besides the average
density for each spin component
\[
n_{\sigma }=\frac{1}{V}\sum_{\mathbf{p}}\left\langle \hat{\psi}_{\mathbf{%
p,\sigma }}^{\dag }\hat{\psi}_{\mathbf{p,\sigma }}\right\rangle ,
\]%
we also need to introduce mean-field parameter associating with the
spin-flip term
\[
\xi =\frac{1}{V}\sum_{\mathbf{p}}\left\langle \hat{\psi}_{\mathbf{p}%
,\uparrow }^{\dag }\hat{\psi}_{\mathbf{p},\downarrow }\right\rangle
\]%
The spin-flip term, $\xi$, is due to the Raman coupling between
different spin states, which breaks the conservation of spin magnetization. In contrast, such a term is absent for an
ordinary spinor Bose gas without Raman coupling.

Up to a constant, the mean-field Hamiltonian in Hartree-Fock
approximation is given by
\begin{eqnarray}
\hat{H}_{\mathrm{HF}} &=&\hat{H}_{0}+\sum_{\mathbf{p}}\Big[g\left(
2n_{\uparrow }+n_{\downarrow }\right) \hat{\psi}_{\mathbf{p,\uparrow }%
}^{\dag }\hat{\psi}_{\mathbf{p,\uparrow }}+g\left( 2n_{\downarrow
}+n_{\uparrow }\right) \hat{\psi}_{\mathbf{p,\downarrow }}^{\dag }\hat{\psi}%
_{\mathbf{p,\downarrow }}  \nonumber  \label{MF-Hamiltonian} \\
&&\qquad \qquad +g\xi \left( \hat{\psi}_{\mathbf{p,\uparrow }}^{\dag }\hat{%
\psi}_{\mathbf{p,\downarrow }}+\hat{\psi}_{\mathbf{p,\downarrow }}^{\dag }%
\hat{\psi}_{\mathbf{p,\uparrow }}\right) \Big]
\end{eqnarray}%
Since we are treating the normal state above $T_{\mathrm{c}}$, spin
are always unpolarized with $n_{\uparrow }=n_{\downarrow }=n/2$. The
mean field Hamiltonian in (\ref{MF-Hamiltonian}) has the same
structure as the single particle Hamiltonian, except that the Raman
coupling is modified as $\Omega _{\mathrm{eff}}=\Omega +2g\xi $ and
the energy zero-point is shifted by a constant. In this sense, we
can define two dressed helicity branches, with dispersions given by
\begin{equation}
\varepsilon _{\mathbf{p},\pm }=\frac{1}{2m}\left( \mathbf{p}%
^{2}+k_{r}^{2}\right) +{\frac{3}{2}}gn\pm \sqrt{\left( k_{r}p_{x}/m\right)
^{2}+\Omega _{\mathrm{eff}}^{2}/4}  \label{normal_dispersion}
\end{equation}%
Here the mean-field parameter $\xi $ should be solved
self-consistently in combination with
\begin{equation}
\xi =\frac{1}{V}\sum_{\mathbf{p}}\sin \theta _{\mathbf{p}}\cos \theta _{%
\mathbf{p}}\left( n_{\mathbf{p,+}}-n_{\mathbf{p,-}}\right)  \label{xi}
\end{equation}%
where $\theta _{\mathbf{p}}$ is given by Eq. (\ref{sincos}) with the
bare Raman coupling replaced by $\Omega _{\mathrm{eff}}$, and
$n_{\mathbf{p,\pm }}$ are Bose distribution functions with the
dispersion $\varepsilon _{\mathbf{p,\pm }}$ given by Eq.
(\ref{normal_dispersion}). The transition temperature $T_c$ is then determined by interaction modified dispersions. Due to the non-trivial momentum-dependence of dispersions (\ref{normal_dispersion}), the interactions induce a shift of $T_c$ from non-interacting Bose gas.

Since the occupation of lower helicity branch $n_{\mathbf{p,-}}$ is
always larger than $n_{\mathbf{p,+}}$, $\xi$ is always negative and
hence the effective Raman coupling strength $\Omega_{\mathrm{eff}}$
is decreased to a smaller value. Based on the non-monotonic behavior
of non-interacting $T_c$ discussed previously, the minimum of the
condensation temperature is shifted by the interaction to a larger
$\Omega$, as shown in Fig. \ref{Tc_uniform} (a). Near the minimum,
the correction of $T_c$ change
its sign rapidly within a narrow region of $\Omega/E_r$. From Fig. \ref{Tc_uniform}%
(b), one can clearly see that the transition temperature is enhanced for $%
\Omega \lesssim 4E_{\mathrm{r}}$ and is suppressed for $\Omega \gtrsim 4E_{%
\mathrm{r}}$, and in between the correction $\Delta T_c=T_c-T^0_c$ shows a most
profound effect around the $T_c$ minimum.

\subsection{Interaction Shift of Transition Temperature for Trapped Gases}

Finally we come to discuss the most realistic case where both interaction
 and trap effect are taken into account. We use the Hatree-Fock
mean-field theory to include interaction effect and use the semi-classical
approximation to include trap effect.

Within the Hatree-Fock and semi-classical approximation,
$T_{\mathrm{c}}$ is determined when $\mu (T)$ reaches the minimum of
single particle spectrum, and $\mu (T)$ is determined by particle
number conservation
\begin{equation}
N=\int d^{3}rn\left( \mathbf{r},T,\mu \right)
\end{equation}%
Here the local density $n\left( \mathbf{r},T,\mu \right) $ is given by
\begin{equation}
n\left( \mathbf{r}\right) =\int \frac{d^{3}k}{\left( 2\pi \right) ^{3}}%
\left\{ \frac{1}{e^{\left[ \varepsilon _{+,\mathbf{k}}\left( \mathbf{r}%
\right) -\mu \right] /T}-1}+\frac{1}{e^{\left[ \varepsilon _{-,\mathbf{k}%
}\left( \mathbf{r}\right) -\mu \right] /T}-1}\right\}
\end{equation}%
in which
\begin{equation}
\varepsilon _{\pm ,\mathbf{k}}\left( \mathbf{r}\right) =\frac{1}{2m}\left(
\mathbf{k}^{2}+k_{r}^{2}\right) \pm \sqrt{\left( k_{r}k_{x}/m\right)
^{2}+\Omega _{\mathrm{eff}}^{2}\left( \mathbf{r}\right) /4}+V_{\mathrm{eff}%
}\left( \mathbf{r}\right) ,  \label{E_sp_trap}
\end{equation}%
and
\begin{equation}
V_{\mathrm{eff}}\left( \mathbf{r}\right) =V\left( \mathbf{r}\right) +(3/2)gn\left( \mathbf{r}\right) .
\end{equation}%
Similar as above, $\Omega _{\mathrm{eff}}$ is the renormalized coupling as
\begin{equation}
\Omega _{\mathrm{eff}}\left( \mathbf{r}\right) =\Omega +2 g\xi \left(
\mathbf{r}\right) .
\end{equation}%
which also needs to be determined self-consistently as Eq.\ref{xi}.

The numerical solution gives the interaction shift of
$T_{\mathrm{c}}$ inside a harmonic trap, as shown in Fig.
\ref{Tc_trap}. Here we notice that
the interaction effect always gives a negative shift of $\Delta T_{\mathrm{c}%
}$, in contrast to the uniform case where $\Delta T_{\mathrm{c}}$
can be either positive or negative. This is because the presence of
the repulsive interaction reduces the central density in the trap, leading to a reduced $%
T_{\mathrm{c}}$. This effect dominates over the shift of $\Omega _{\mathrm{%
eff}}$ for sufficient large particle number. In Fig. \ref{Tc_trap} we also
plot $|\Delta T_{\mathrm{c}}|/T_{\mathrm{c}}^{0}$ as a function of $\Omega _{%
\mathrm{c}}$, and find the relative shift of $T_{\mathrm{c}}$
reaches a maximum around $\Omega =4E_{\mathrm{r}}$. This can be
qualitatively understood from the effective mass approximation.
Using $m^*$, one can define an effective scattering
length $a_{s}^*=a_s(m^*/m)$ and harmonic length $a_{\mathrm{ho}}^*=a_{\mathrm{ho}} (m/m^*)^{1/4}$; therefore the shift of $T_{%
\mathrm{c}}$ in the harmonic trap\cite{Giorgini} is $\Delta T_{\mathrm{c}}/T_{\mathrm{c}}^{0}=-1.32\frac{%
a_{\mathrm{s}}^*}{a_{\mathrm{ho}}^*}N^{1/6}\propto (m^*/m)^{5/4}$.
As we have shown in Eq.\ref{meff}, $m^*/m$ is maximally enhanced at $\Omega =4E_{\mathrm{r}}$, so the interaction has the maximum effect to the shift of $T_c$ at this point.

\section{Conclusion}

In this paper, we investigate the properties of Bose gases with
Raman-induced spin-orbit coupling. Our main results are summarized as follows.

(1)The presence of the SO coupling modifies the single particle spectrum and thus
the single particle DoS. At $\Omega =4E_{\mathrm{r}}$, the low energy DoS reaches a maximum and the effective mass is maximally enhanced.
The direct consequences include the vanishment of sound velocity at $\Omega
=4E_{\mathrm{r}}$, and the non-monotonic behavior of condensate depletion, Lee-Huang-Yang correction of ground-state energy, and the transition temperature of a non-interacting Bose-Einstein condensate.

(2) The presence of the SO coupling breaks the Galilean invariance. As a result, the critical dragging and flowing velocity, respectively defined in the rest frame of the condensate and the impurity, are no longer identical.

(3) In the presence of the SO coupling, a roton minimum will appear in the excitation spectrum in the regime of $\Omega<4E_r$. As a result, the thermal depletion of the condensate can possess an opposite magnetization with the quantum depletion. Moreover, the critical dragging velocity exhibits asymmetry along different directions.

(4)  In the presence of the SO coupling,  the
interactions shift BEC transition temperature $T_c$ even at a
Hartree-Fock level. In both homogeneous and trapped systems, the interaction shift of $T_c$ shows maximum around $\Omega =4E_{\mathrm{r}}$, where the interactions take the largest effect due to the enhancement of DoS and the effective mass.

In conclusion, we have shown that the Bose gases with Raman-induced SO coupling can exhibit a number of non-trivial properties, as summarized above. The results revealed here can be directly verified in the current cold atom experiments using laser-induced gauge field.

\end{document}